\documentclass[aps,twocolumn,prx,preprintnumbers,showpacs,amsmath,amssymb,superscriptaddress]{revtex4-1}
\usepackage[pdftex, colorlinks, citecolor=blue]{hyperref}   
\usepackage{graphicx}
\usepackage{dcolumn}
\usepackage{threeparttable}
\usepackage{multirow}
\usepackage{booktabs}
\usepackage{txfonts}
\usepackage{xcolor}
\usepackage{bm}
\usepackage{amssymb}
\usepackage{amsmath}
\usepackage{latexsym}
\usepackage{epsfig}
\usepackage{amsbsy}
\usepackage{array}
\usepackage{tabularx}
\usepackage{esvect} 
\usepackage{extarrows}
\usepackage{float}
\usepackage[british]{babel}

\begin{document}

\title{First-principles study on the electronic structure of Pb$_{10-x}$Cu$_x$(PO$_4$)$_6$O ($x$=0, 1) }

\author{Junwen Lai}
\thanks{These authors contribute equally to this work.}
\affiliation{%
Shenyang National Laboratory for Materials Science, Institute of Metal Research, Chinese Academy of Sciences, 110016 Shenyang, China.
}%
\affiliation{%
School of Materials Science and Engineering, University of Science and Technology of China, 110016 Shenyang, China.
}%

\author{Jiangxu Li}
\thanks{These authors contribute equally to this work.}
\affiliation{%
Shenyang National Laboratory for Materials Science, Institute of Metal Research, Chinese Academy of Sciences, 110016 Shenyang, China.
}%

\author{Peitao Liu}%
\email{ptliu@imr.ac.cn}
\affiliation{%
Shenyang National Laboratory for Materials Science, Institute of Metal Research, Chinese Academy of Sciences, 110016 Shenyang, China.
}%

\author{Yan Sun}
\email{sunyan@imr.ac.cn}
\affiliation{%
Shenyang National Laboratory for Materials Science, Institute of Metal Research, Chinese Academy of Sciences, 110016 Shenyang, China.
}%

\author{Xing-Qiu Chen}%
\affiliation{%
Shenyang National Laboratory for Materials Science, Institute of Metal Research, Chinese Academy of Sciences, 110016 Shenyang, China.
}%

\begin{abstract}
Recently, Lee \emph{et al.} reported the experimental discovery of room-temperature ambient-pressure superconductivity in a Cu-doped lead-apatite (LK-99) (arXiv:2307.12008, arXiv:2307.12037). Remarkably, the superconductivity persists up to 400 K at ambient pressure. Despite strong experimental evidence, the electronic structure of LK-99 has not yet been studied. Here, we investigate the electronic structures of LK-99 and its parent compound using first-principles calculations, aiming to elucidate the doping effects of Cu. Our results reveal that the parent compound Pb$_{10}$(PO$_4$)$_6$O is an insulator, while Cu doping induces an insulator-metal transition and thus volume contraction. The band structures of LK-99 around the Fermi level are featured by a half-filled flat band and a fully-occupied flat band. These two flat bands arise from both the $2p$ orbitals of $1/4$-occupied O atoms and the hybridization of the $3d$ orbitals of Cu with the $2p$ orbitals of its nearest-neighboring O atoms. Interestingly, we observe four van Hove singularities on these two flat bands. Furthermore, we show that the flat band structures can be tuned by including electronic correlation effects or by doping different elements. We find that among the considered doping elements (Ni, Cu, Zn, Ag, and Au), both Ni and Zn doping result in the gap opening, whereas Au exhibits doping effects more similar to Cu than Ag. Our work provides a foundation for future studies on the role of unique electronic structures of LK-99 in superconductivity.
\end{abstract}

\maketitle

\section{Introduction}

Superconductors are highly valued in various fields due to their unique characteristics of zero resistance and perfect diamagnetism~\cite{book_Philippe}.
This has led experimental and theoretical scientists to tirelessly search for high-$T_c$ superconductors, resulting in numerous breakthroughs in the field, such as the non-hydrogen BCS superconductor MgB$_2$ ($T_c$=39 K, ambient pressure)~\cite{Nagamatsu2001},
the Cu-based unconventional superconductor Hg-Ba-Ca-Cu-O system ($T_c$=133 K, ambient pressure)~\cite{Schilling1993},
the Fe-based unconventional superconductor single-layer FeSe film ($T_c$=77 K, ambient pressure)~\cite{FeSe2012CPL},
and the Ni-based unconventional superconductor La$_3$Ni$_2$O$_7$ ($T_c$=80 K, 14.0$\sim$43.5 GPa)~\cite{Sun2023}.
Owing to Neil Ashcroft's seminal insights that
metallic hydrogen under sufficiently high pressures and
hydrides at lower pressures by chemical precompression may exhibit room-temperature superconductivity~\cite{PhysRevLett.21.1748,PhysRevLett.92.187002},
the search has turned to the superhydrides~\cite{Ma2014, Duan2014, Drozdov2015, Drozdov2019,SomayazuluPRL2019,
SEMENOK202036, YH6_2021, SniderPRL2021, DuanPRL_2022, MaPRL2023_LaBeH8, SEMENOK2020100808, Zhang2021MRE,
MaYM2019PRL,PhysRevB.102.014516,doi:10.1021/acs.jpcc.1c10976,10.3389/femat.2022.837651, PhysRevB.107.L060501},
leading to the so-called ``hydride rush"~\cite{Ma2020MRE,FLORESLIVAS20201,Lilia_2022}.
Despite extensive research, the $T_c$ remains below the room temperature, or the required pressure is too high,
which greatly hinders practical applications.
While the near-ambient superconductivity has been recently claimed in a nitrogen-doped lutetium hydride~\cite{LuNH_Nature},
its validity remains a subject of ongoing debate~\cite{Ming2023,arXiv:2303.17587,salke2023evidence,peng2023near}.

Rather than focusing on hydride superconductors, a groundbreaking discovery was recently reported by Lee \emph{et al.}~\cite{Lee2023_1,Lee2023_2}, claiming the realization of the first room-temperature ambient-pressure superconductor in a Cu-doped lead-apatite (LK-99). The reported $T_c$ is as high as 400 K at ambient pressure, and relevant experimental characterizations, such as resistivity, Meissner effect, and I-V characteristics, were performed~\cite{Lee2023_2}. Moreover, the authors provided an exhaustive sample synthesis method, which would enable other experimental groups to readily reproduce their findings. Despite the strong experimental evidence, several important questions remain open:
What are the electronic structures of the LK-99 and its parent compound,
what is the role played by Cu,
and is it possible to achieve similar effects through doping other elements?

In this work, we set out to address these issues using first-principle density functional theory (DFT) calculations.
We find that the parent compound Pb$_{10}$(PO$_4$)$_6$O of LK-99 is an insulator exhibiting two flat bands below the Fermi level.
Doping Cu induces an insulator-metal transition, which is
featured by a half-filled flat band and a fully-occupied flat band around the the Fermi level.
These two flat bands come from both the $2p$ orbitals of $1/4$-occupied O2 atoms and
the hybridization of the $3d$ orbitals of Cu with the $2p$ orbitals of its nearest-neighboring O1 atoms,
necessitating a minimal two-band low-energy effective model for describing LK-99.
In addition, we observe four van Hove singularities (VHSs) on these two flat bands
stemming from the saddle dispersions at the M and L points in the Brillouin zone.
We show that the flat band structures can be tuned by electronic correlations
or by doping different elements. Among the considered doping elements (Ni, Cu, Zn, Ag, and Au),
both Ni and Zn doping open the band gap, whereas Au exhibits doping effects more similar to Cu than Ag.

\begin{figure*}
\begin{center}
\includegraphics[width=0.9\textwidth, clip]{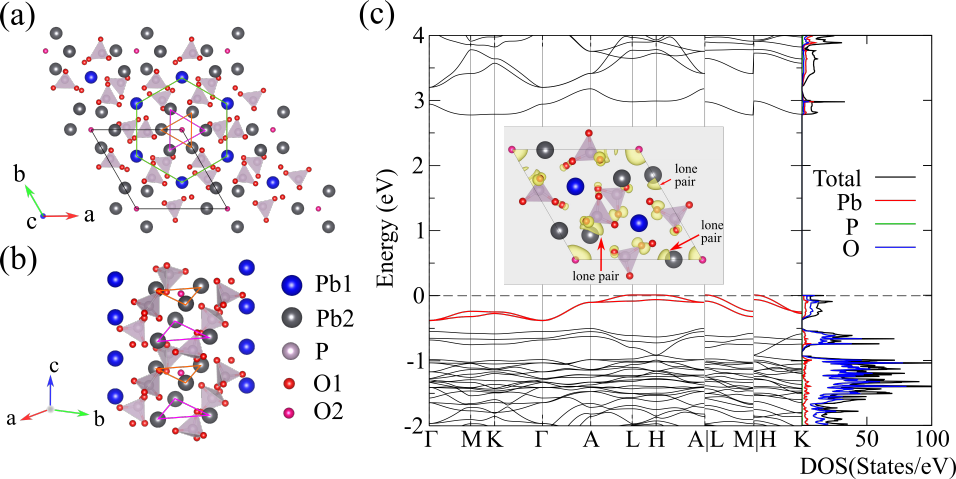}
\end{center}
\caption{(a) Top view of the crystal structure of lead-apatite with a chemical formula of Pb$_{10}$(PO$_4$)$_6$O.
The black lines indicate the unit cell that is periodically replicated.
The Pb1 atoms form a hexagon marked by green lines.
The Pb2 atoms form two oppositely shaped triangles marked by pink and orange lines.
The O2 atoms are $1/4$ occupied.
The O1 atoms denote other O atoms (excluding O2) including the ones forming PO$_4$ units.
(b) Side view of  the crystal structure, highlighting a cylindrical column centered at O2 atoms.
(c) The calculated electronic band structure and density of states.
The inset shows the isosurfaces of calculated charge densities (isovalue=0.006 e/$\AA^3$)
for the two flat bands (in red color) below the Fermi level.
}
\label{Fig1_parent}
\end{figure*}

\section{Computational details}
The first-principles calculations were performed
using the Vienna \emph{ab initio} simulation package (VASP)~\cite{PhysRevB.47.558, PhysRevB.54.11169}.
The plane-wave cutoff was chosen to be 520 eV.
A $\Gamma$-centered $k$-point grid with a $k$-spacing of 0.03 $2\pi/\AA$ was used for structural relaxations
and the $k$-spacing was reduced to 0.02 $2\pi/\AA$ for density of states (DOSs) calculations,
The electronic interactions were described using the Perdew-Burke-Ernzerhof (PBE) functional~\cite{PhysRevLett.77.3865}.
The VASP recommended projector augmented wave pseudopotentials~\cite{PhysRevB.50.17953,PhysRevB.59.1758} were employed.
The tetrahedron method with Bl\"{o}chl corrections was used for DOSs calculations,
whereas the Gaussian smearing method with a smearing width of 0.05 eV was used for other calculations.
The convergence criteria for the total energy and ionic forces were set to 10$^{-6}$ eV and 0.01 eV/$\AA$, respectively.
The electronic correlation effects were described using the DFT+$U$ method based on the Dudarev scheme~\cite{PhysRevB.57.1505}.

\section{Results and discussions}

\subsection{Structural and electronic properties of lead-apatite}

Figure~\ref{Fig1_parent}(a) shows the crystal structure of lead-apatite with a chemical formula of Pb$_{10}$(PO$_4$)$_6$O.
It is the parent compound of LK-99 with a hexagonal ($P6_3/m$,  space group 176)  lattice.
The calculated lattice parameters are $a$=10.024 $\AA$ and $c$=7.482 $\AA$,
in good agreement with the experimental values ($a$=9.865 $\AA$ and $c$=7.431 $\AA$)~\cite{ExptStr2003}.
The predicted larger lattice parameters arise from the overestimation of the employed PBE functional.
In the lead-apatite, there exists two symmetry-inequivalent Pb atoms, named Pb1 and Pb2.
The the Pb1 atoms form a hexagon, whereas the Pb2 atoms form two oppositely shaped triangles [see Fig.~\ref{Fig1_parent}(a)].
Along the $c$ axis, the Pb2 atoms along with the surrounding PO$_4$ units form a cylindrical column centered at O2 atoms [see Fig.~\ref{Fig1_parent}(b)].
We note that the O2 atoms are 1/4 occupied. To simulate this, we removed three out of four O2 atoms in calculations.
The employed structure is provided in the Supplementary Material~\cite{SM}.

The calculated electronic band structure and DOSs of lead-apatite are shown in Fig.~\ref{Fig1_parent}(c).
It is obvious that the lead-apatite is a nonmagnetic insulator with a PBE-predicted indirect gap of 2.77 eV.
The obtained insulating nature of lead-apatite is consistent with the experimental observations~\cite{Lee2023_1,Lee2023_2}.
It should be stressed that the 1/4 occupation of O2 atoms is important to open the band gap.
Full occupation of O2 atoms results in a metal.
DOSs show that the valence bands in the energy region of [$-$2 eV, 0 eV] are contributed by the dominate O-$2p$ states and minor Pb-($6s+6p$) states.
Notably, two flat bands right below the Fermi level are observed.
The calculated charge densities reveal that they are mainly from the $2p$-states of O2 atoms,
the $2p$-states of the nearest-neighboring O1 atoms of Pb1 atoms,
and the stereochemically active $6s^2$ lone pairs of Pb2 atoms [insert of Fig.~\ref{Fig1_parent}(c)].
The existence of the Pb2-$6s^2$ lone pairs leads to asymmetrical arrangements of neighboring six O atoms.

\subsection{Electronic structure of  LK-99}

\begin{figure*}
\begin{center}
\includegraphics[width=0.93\textwidth, clip]{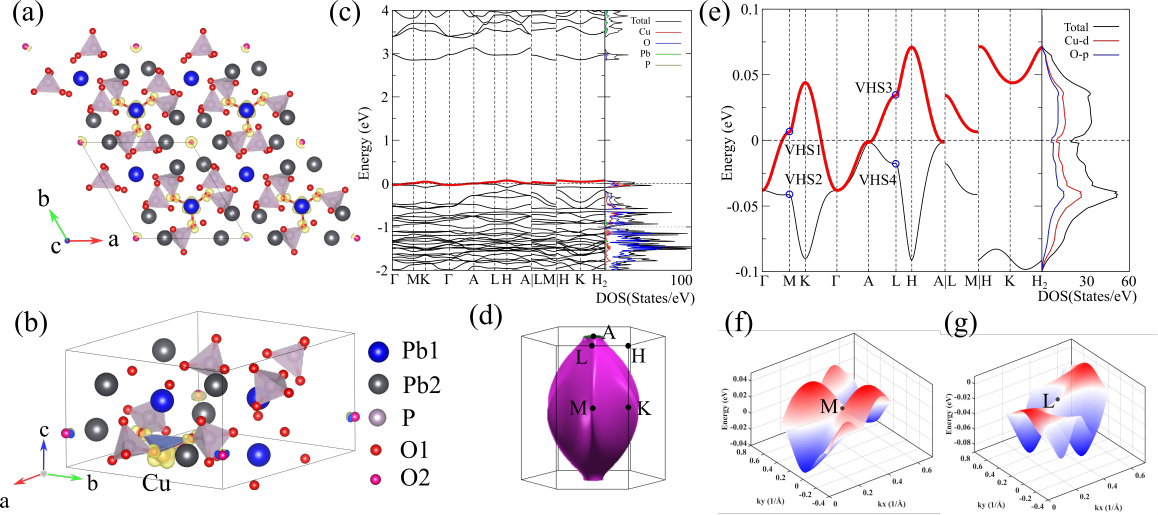}
\end{center}
\caption{(a, b) Top and side views of the crystal structure of LK-99 with a chemical formula of Pb$_{9}$Cu(PO$_4$)$_6$O.
Here, the isosurfaces of calculated charge densities (isovalue=0.011 e/$\AA^3$)
for the two flat bands around the Fermi level [red lines in (c)] are superimposed.
(c) The calculated electronic band structure and density of states.
(d) The computed Fermi surface.
(e) Zoom-in of (c), highlighting the half-filled flat band and four VHSs marked by VHS1 and VHS2 at M and VHS3 and VHS4 at L.
(f, g) The three-dimensional visualizations of the half-filled flat band around the M and L points, respectively, highlighting the saddle dispersions around VHS1 and VHS3.
Note that all the results here were obtained using non-spin-polarized PBE calculations.
}
\label{LK-99}
\end{figure*}

Having analyzed structural and electronic properties of lead-apatite, we turn to LK-99.
Let us start our discussion with the high-temperature nonmagnetic solution without considering on-site Coulomb interactions on the Cu-$3d$ shell.
In experiment, the LK-99 exhibits a chemical formula of Pb$_{10-x}$Cu$_x$(PO$_4$)$_6$O with 0.9$<$x$<$1.1~\cite{Lee2023_2}.
Furthermore, it was found that the Cu atoms occupy the Pb1 positions in the crystal structure~\cite{Lee2023_2}.
For simplicity, we only considered the case with $x$=1, which can be achieved by substituting a Pb1 atom with a Cu atom, resulting in Pb$_{9}$Cu(PO$_4$)$_6$O.
We note that Cu doping induces symmetry breaking, resulting in four possible configurations (see Supplementary Material Fig.~S1~\cite{SM}).
These configurations have total energies that are very close to each other ($<$5 meV/atom), but in this study, we focus on the configuration with the lowest energy, whose crystal structure is shown in Figs.\ref{LK-99}(a) and (b) and provided in the Supplementary Material~\cite{SM}.
The calculated lattice parameters of LK-99 ($a$=9.931 $\AA$ and $c$=7.411 $\AA$)
are in good agreement with the experimental values ($a$=9.843 $\AA$ and $c$=7.428 $\AA$)~\cite{Lee2023_2}.
As compared to the non-doped lead-apatite, Cu doping results in a 2.9\% reduction of the system volume (Table~\ref{Table1})
due to the insulator-metal transition, which was argued as one of the key factors
contributing to the realization of room-temperature superconductivity in LK-99~\cite{Lee2023_2}.

\begin{table}
\caption{The PBE predicted lattice parameters $a$ and $c$ (in $\AA$), volume (in $\AA^3$), and band gap (in eV)
of Pb$_{9}$M(PO$_4$)$_6$O where M indicates the elements listed below.
Note that M=Pb corresponds to the parent compound of LK-99.
The available experimental data are given for comparison.
}
\begin{ruledtabular}
\begin{tabular}{ccccc}
M   & $a$  &$c$ & Volume & Band gap \\
\hline
Pb & 10.024 & 7.482  & 651.02  & 2.77   \\
Expt.\cite{ExptStr2003}  & 9.865 & 7.431  & 626.25  & Insulator   \\
\hline
Cu  & 9.931 & 7.411  &  632.94  & Metal \\
Expt.\cite{Lee2023_2}  & 9.843 & 7.428  & 623.24 & Metal   \\
\hline
Ni   & 9.815 & 7.386  & 616.29   & 0.82   \\
Zn   & 9.833 & 7.384   & 618.34   & 2.94  \\
Ag  & 9.941 & 7.382  & 631.69   & Metal \\
Au  & 9.928 & 7.421  & 633.48   & Metal \\
\end{tabular}
\end{ruledtabular}
\label{Table1}
\end{table}

\begin{figure*}
\begin{center}
\includegraphics[width=0.88\textwidth, clip]{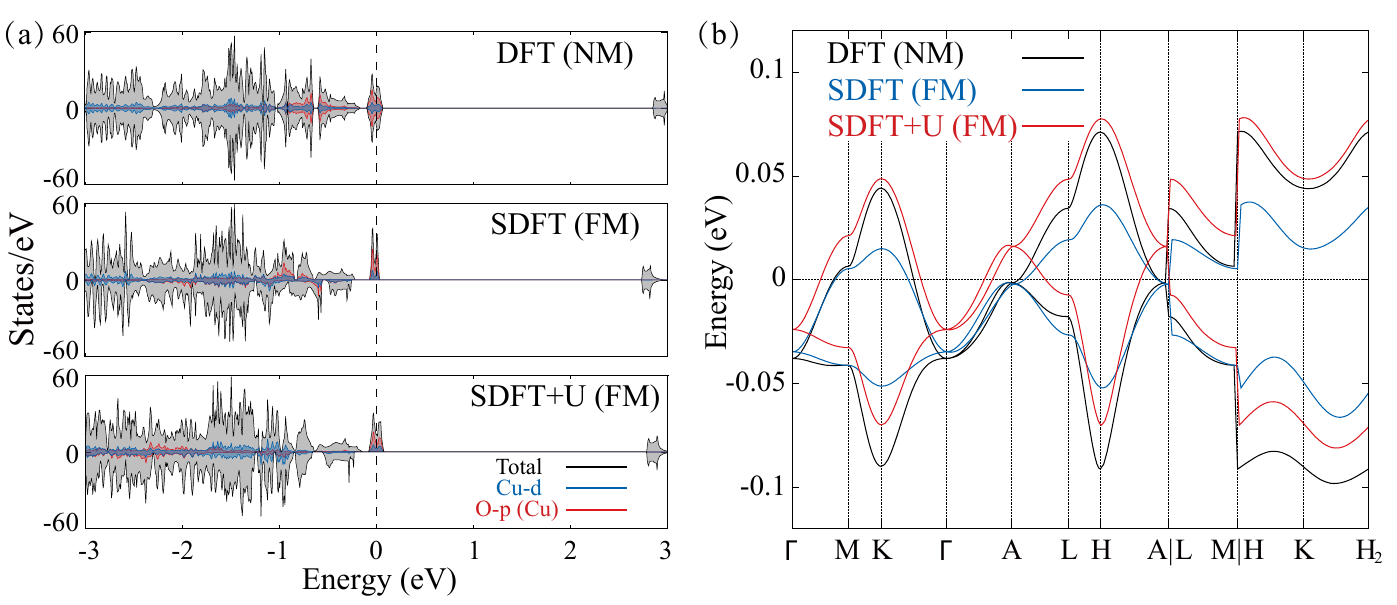}
\end{center}
\caption{Effects of spin polarizations and electronic correlations on the (a) DOSs and (b) band structures of LK-99 around the Fermi level.
SDFT denotes the spin-polarized DFT. NM and FM represent the nonmagnetic and ferromagnetic solutions, respectively. 
}
\label{Fig3_correlation_effect}
\end{figure*}

Figures~\ref{LK-99}(c) and (e) show the calculated band structure and DOSs of LK-99.
Clearly, after Cu doping, the LK-99 becomes metallic,
with just one half-filled very flat band (dispersion$<$0.15 eV) crossing the Fermi level.
The Fermi surface looks like a rugby and consists of both electron and hole pockets [Fig.~\ref{LK-99}(d)].
Below this half-filled flat band there exists another flat band that is fully occupied.
We note that in comparison with other correlated $d$-band superconductors, 
these two low-energy bands of LK-99 are much flatter. 
Moreover, unlike the widely studied one-band Hubbard model applicable to cuprates~\cite{Gull2015,doi:10.1126/science.aal5304},
here a minimal two-band low-energy effective model should be used to describe the physics of LK-99,
since the two flat bands are contributed by both the $2p$ states of O2 atoms
and the hybridization of the Cu-$3d$ states
with the $2p$ states of its nearest-neighboring O1 atoms [see Figs.~\ref{LK-99}(a) and (b)].
It is interesting to note that the conduction electrons are confined in the layer containing doped Cu atoms,
whereas other layers without doped Cu atoms seem insulating.
In addition, the PO$_4$ units surrounding the cylindrical column formed by Pb2 atoms also exhibit insulating characteristics,
leading to a one-dimensional-like conduction channel along the $c$ axis mediated by the 1/4-occupied O2 atoms.
More interestingly, we observed four VHSs on these two flat bands,
originating from the saddle dispersions at the M and L points in the Brillouin zone [see Figs.~\ref{LK-99}(e), (f) and (g)].
This indicates that the electronic properties are fragile in response to structural distortions at low temperatures.
The existence of these unique flat bands may be connected with the experimentally
observed remarkable superconducting properties of  LK-99~\cite{Lee2023_1,Lee2023_2}, which worths further investigation.

Before closing this section, we would like to briefly discuss the effects of spin polarizations and electronic correlations on the electronic structures.
Figure~\ref{Fig3_correlation_effect} compares the DOSs and two flat bands around the Fermi level
obtained from nonmagnetic DFT calculations, spin-polarized DFT calculations, and  spin-polarized DFT+$U$ calculations.
It was found that considering spin polarizations lowers the total energy by 2.4 meV/atom,
leading to a ferromagnetic metal at 0 K with a magnetic moment of 0.57 $\mu_{\rm B}$/Cu
and 0.06 $\mu_{\rm B}$ on its nearest-neighboring O1 atoms due to orbitals hybridization.
Compared to the nonmagnetic solution, the two bands around the Fermi level obtained from the spin-polarized calculations remain flat.
However, they are now fully spin-polarized [see the middle panel of Fig.~\ref{Fig3_correlation_effect}(a)].
The crystal field and projected DOSs analyses suggest that they are dominated by the half-filled Cu-$d_{\rm yz}/d_{\rm xz}$ orbitals.
The same holds when including the Hubbard interaction ($U$= 4 eV~\cite{PhysRevB.73.195107}) on the Cu-$d$ shell,
which also favors the ferromagnetic solution  [see the bottom panel of Fig.~\ref{Fig3_correlation_effect}(a)]. 
The SDFT+$U$ calculated magnetic moments are 0.49 $\mu_{\rm B}$ and 0.08 $\mu_{\rm B}$
for the Cu and  its nearest-neighboring O1 atoms, respectively.
In general, the two flat bands around the Fermi level are quite similar for the three cases.
The noticeable difference lies in the bandwidth. 
One can see from Fig.~\ref{Fig3_correlation_effect}(b) that 
considering the spin polarizations slightly narrows the bandwidth of the two flat bands [compare blue and black curves in Fig.~\ref{Fig3_correlation_effect}(b)],
whereas including the electronic correlations marginally increases the bandwidth [compare red and blue curves in Fig.~\ref{Fig3_correlation_effect}(b)].
This is somewhat unusual as the Hubbard interactions typically localize electrons and thus decrease the bandwidth.
These findings suggest that electronic correlations can fine-tune the flat band structures around the Fermi level
and therefore affect the superconducting properties~\cite{PhysRevB.83.220503,PhysRevB.102.201112}.

\subsection{Electronic structure tuning via different elements doping}

\begin{figure}
\begin{center}
\includegraphics[width=0.49\textwidth, clip]{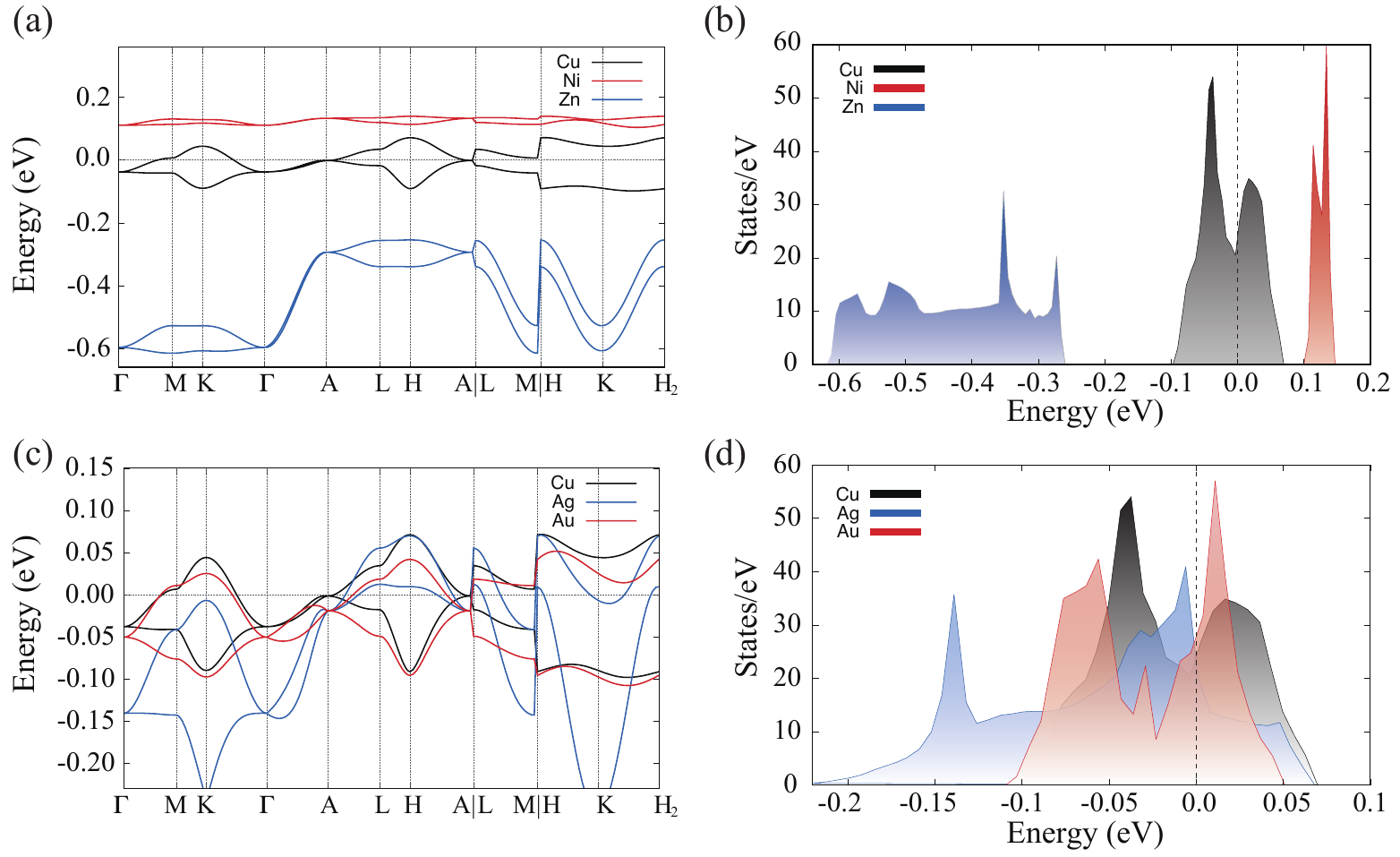}
\end{center}
\caption{Effects of different elements doping on the (a, c) band structures and (b, d) total density of states of LK-99 around the Fermi level.
Note that all the results here were obtained using non-spin-polarized PBE calculations without taking into account on-site Coulomb interactions on the $d$ shell.
}
\label{Fig4_doping_diffelements}
\end{figure}

As previously introduced, LK-99 exhibits novel electronic structures around the Fermi level that may play a crucial role in forming its superconducting state.
A natural question arises as to whether these novel electronic structures are a result of the specific Cu doping,
or in other words, whether alternative elements exist having similar doping effects to Cu.
To address this question, we conducted first-principles band structure calculations using different doping elements.
Specifically, we considered four elements: Ni, Zn, Ag, and Au. Ni and Zn are located on the left and right sides of Cu, respectively, in the periodic table,
and we chose them to examine the effects of electron filling.
Ag and Au belong to the same group as Cu, and we included them to investigate the effects of volume.

Figures~\ref{Fig4_doping_diffelements}(a) and (b) show that both Ni and Zn doping result in gap opening.
This can be understood from the electron filling, since both elements lead to the even number of valence electrons.
Turning to the Ag and Au doping cases, one can see from Figs.~\ref{Fig4_doping_diffelements}(c) and (d)
that the two flat bands around the Fermi level persist due to their similar electronic configurations as Cu (note the small energy scale).
However, we find that the doping effects of Au are more similar to Cu than those of Ag.
This observation may be explained by the closer volumes between Au-doped and Cu-doped lead-apatites (see Table~\ref{Table1}).
It is worth mentioning that, compared to Cu, Au doping further narrows the two flat bands,
brings the VHS1 at the M point and VHS3 at the L point closer to the Fermi level,
and results in larger DOSs around the Fermi level [see Figs.~\ref{Fig4_doping_diffelements}(c) and (d)].
These results suggest that Au-doped lead-apatite may also exhibit similar superconducting properties as the Cu-doped one.

\section{Conclusions}

In conclusion, we have performed a thorough first-principles electronic structure analysis of both LK-99 and its parent compound.
Our results demonstrate that the parent compound without doping is an insulator with a large band gap.
Cu doping closes the band gap, resulting in an insulator-metal transition and volume contraction.
The band structures of LK-99 exhibit novel characteristics around the Fermi level,
which are manifested by the presence of a half-filled flat band and a fully-occupied flat band
with two VHSs each.
These two flat bands arise from both the $2p$ orbitals of $1/4$-occupied O2 atoms
and the hybridization of the $3d$ orbitals of Cu with the $2p$ orbitals of its nearest-neighboring O1 atoms,
necessitating a minimal two-band low-energy effective model.
The former $2p$ orbitals of $1/4$-occupied O2 atoms lead to
a one-dimensional-like cylindrical conduction channel along the $c$ axis surrounded by insulating PO$_4$ units,
whereas the latter hybridized Cu-$3d$ orbitals and O1-$2p$ orbitals
give rise to a two-dimensional-like electron charge distribution in the $ab$-plane.
The observed VHSs are robust against electronic correlation effects
and their band energies can be tuned by doping different elements.
We find that Au exhibits doping effects more similar to Cu than Ag.
We anticipate that the distinctive flat-band electronic structures of LK-99 that we observed in this study will aid in identifying the source of its exceptional superconducting properties.

\section{Acknowledgments}
This work was supported by the National Key R\&D Program of China (No.~2021YFB3501503),
the National Natural Science Foundation of China  (Grant No. 52188101),
and Chinese Academy of Sciences (No. ZDRW-CN-2021-2-5).

\bibliographystyle{apsrev4-1}
\bibliography{Reference} 

\end{document}


\title{Supplementary Material to \\
		``First-principles study on the electronic structure of Pb$_{10-x}$Cu$_x$(PO$_4$)$_6$O ($x$=0, 1)"}

	\author{Junwen Lai}
	\thanks{These authors contribute equally to this work.}
	\affiliation{%
		Shenyang National Laboratory for Materials Science, Institute of Metal Research, Chinese Academy of Sciences, 110016 Shenyang, China.
	}%
	\affiliation{%
		School of Materials Science and Engineering, University of Science and Technology of China, 110016 Shenyang, China.
	}%

	\author{Jiangxu Li}
	\thanks{These authors contribute equally to this work.}
	\affiliation{%
		Shenyang National Laboratory for Materials Science, Institute of Metal Research, Chinese Academy of Sciences, 110016 Shenyang, China.
	}%

	\author{Peitao Liu}%
	\email{ptliu@imr.ac.cn}
	\affiliation{%
		Shenyang National Laboratory for Materials Science, Institute of Metal Research, Chinese Academy of Sciences, 110016 Shenyang, China.
	}%
	
	\author{Yan Sun}
	\email{sunyan@imr.ac.cn}
	\affiliation{%
		Shenyang National Laboratory for Materials Science, Institute of Metal Research, Chinese Academy of Sciences, 110016 Shenyang, China.
	}%
	
	\author{Xing-Qiu Chen}%
	\affiliation{%
		Shenyang National Laboratory for Materials Science, Institute of Metal Research, Chinese Academy of Sciences, 110016 Shenyang, China.
	}%

	\maketitle

	\begin{figure}
		\begin{center}
			\includegraphics[width=0.80\textwidth, clip]{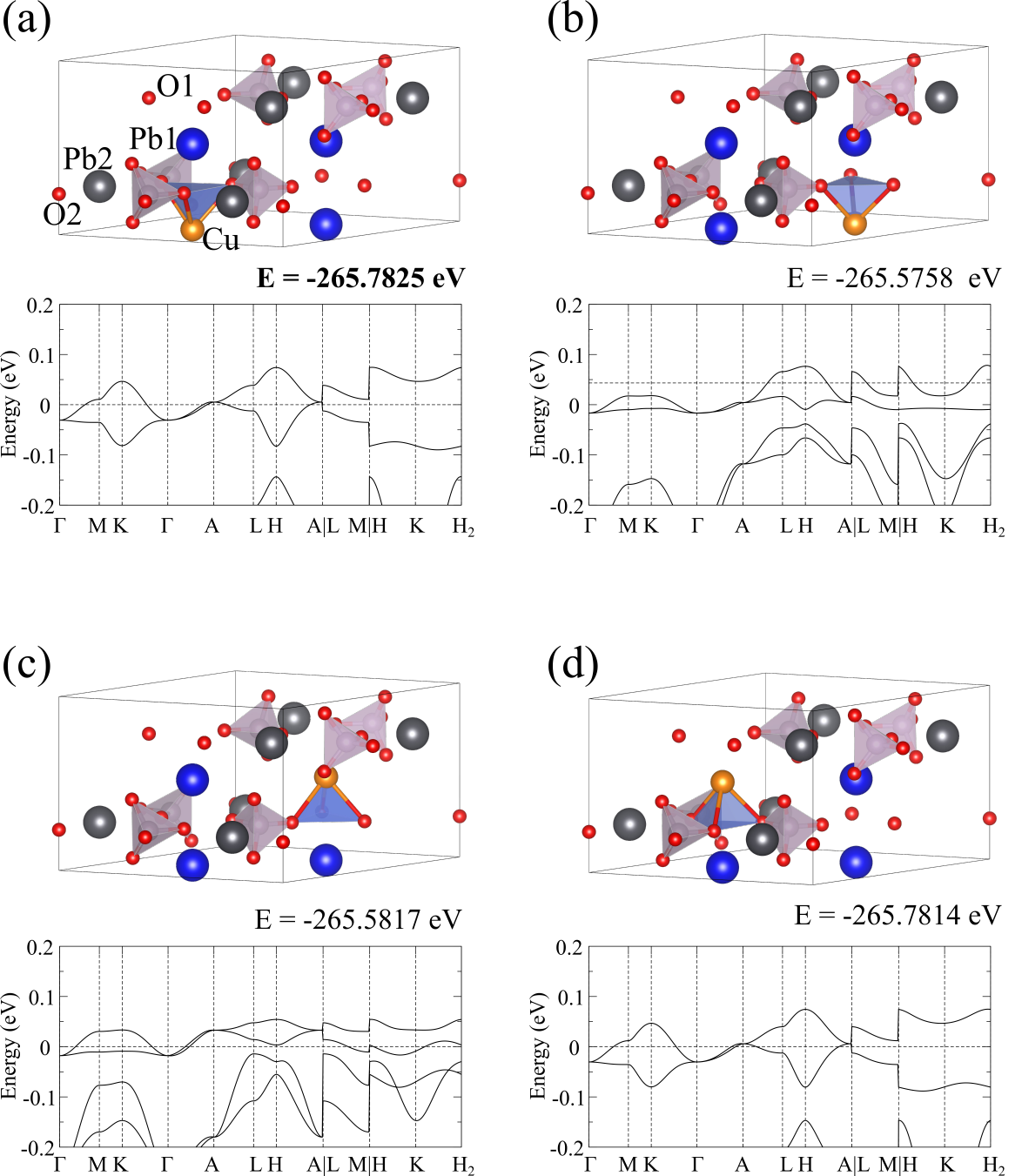}
		\end{center}
		\caption{Crystal and band structures of four possible lead-apatite configurations with a Cu atom substituting a Pb1 site.
From the band structures, it seems that (a) and (d) are symmetry-equivalent configurations,
and (b) and (c) are symmetry-equivalent configurations.
In any case, there exist two flat bands crossing the Fermi level.
The energies difference between (a) and (b) are small ($<$ 5 meV/atom).
The adopted Cu-doped structure model [i.e., Fig.~2(b) of the main text] corresponds to the configuration (a) here.
Note that all the results here were obtained from non-spin-polarized PBE calculations .
		}
		\label{FigS1}
	\end{figure}


%

\begin{verbatim}
Pb10(PO4)6O
   1.00000000000000
    10.0239401044820102    0.0000006121514461    0.0000000000000000
    -5.0119705824320677    8.6809864704190467    0.0000000000000000
     0.0000000000000000    0.0000000000000000    7.4815041019845081
   Pb   P    O
    10     6    25
Direct
  0.9963538867363795  0.7711561244778835  0.2443294850732514
  0.0050062123768200  0.2643090859692521  0.7464431433615388
  0.2288438755221165  0.2251977622584960  0.2443294850732514
  0.7356909140307479  0.7406971274075715  0.7464431433615388
  0.7748022527415088  0.0036461272636217  0.2443294850732514
  0.2593028875924333  0.9949938016231812  0.7464431433615388
  0.6666666870000029  0.3333333429999996  0.0042634124695411
  0.3333333129999971  0.6666666269999979  0.0072229018756005
  0.3333333129999971  0.6666666269999979  0.4870401649864249
  0.6666666870000029  0.3333333429999996  0.4951010130162885
  0.6239014908271940  0.5947299148479956  0.2487979403241596
  0.3668872594614285  0.3950708418260263  0.7483756092457483
  0.4052700851520044  0.0291715459791888  0.2487979403241596
  0.6049291581739737  0.9718164476354048  0.7483756092457483
  0.9708284840208137  0.3760985391728084  0.2487979403241596
  0.0281835823645977  0.6331127705385740  0.7483756092457483
  0.5006906503976367  0.6458112695516860  0.2472504771707875
  0.4689633991493309  0.3196890005137618  0.7495392298128181
  0.3541887304483140  0.8548793808459507  0.2472504771707875
  0.6803109994862382  0.1492743976355726  0.7495392298128181
  0.1451205891540468  0.4993093206023573  0.2472504771707875
  0.8507255723644249  0.5310365718506631  0.7495392298128181
  0.7292899627055505  0.6547134078302221  0.0796661284193476
  0.2609217265457389  0.3521073372121108  0.9164206285216281
  0.3452865921697779  0.0745765248753330  0.0796661284193476
  0.6478926627878892  0.9088144183336340  0.9164206285216281
  0.9254235041246730  0.2707100672944520  0.0796661284193476
  0.0911856106663720  0.7390783034542636  0.9164206285216281
  0.2583831295645709  0.3475865624117134  0.5824263589771803
  0.7304908854720296  0.6591553131478420  0.4164438897440874
  0.6524134375882866  0.9107965961528564  0.5824263589771803
  0.3408446868521580  0.0713355423241993  0.4164438897440874
  0.0892034328471496  0.7416169004354316  0.5824263589771803
  0.9286644866758067  0.2695091445279658  0.4164438897440874
  0.5445286861139778  0.4147734915236683  0.2515084873369702
  0.4721921906908833  0.5767053663538633  0.7451733587211393
  0.5852264784763292  0.1297551945903166  0.2515084873369702
  0.4232946036461342  0.8954867643370079  0.7451733587211393
  0.8702447754096809  0.4554713138860222  0.2515084873369702
  0.1045132056629896  0.5278078093091167  0.7451733587211393
  0.0000000000000000  0.0000000000000000  0.2339466544173519
\end{verbatim}

\begin{verbatim}
	Pb9Cu(PO4)6O
	1.00000000000000
	9.9306823821051555   -0.0000014818366361    0.0000000000000000
	-4.9653399077953200    8.6002239607356703   -0.0000000000000000
	0.0000000000000000    0.0000000000000000    7.4109945172796605
	Pb   Cu   P    O
	9    1    6    25
	Direct
	0.9986807449642078  0.7699202447216552  0.2474756380877509
	0.9977014382162520  0.2577003216124319  0.7545056960306785
	0.2300797552783452  0.2287605002425670  0.2474756380877509
	0.7422996783875685  0.7400011176038168  0.7545056960306785
	0.7712395147574378  0.0013192690357936  0.2474756380877509
	0.2599988973961879  0.0022985757837491  0.7545056960306785
	0.3333333129999971  0.6666666269999979  0.0102252768912011
	0.3333333129999971  0.6666666269999979  0.4962563424132084
	0.6666666870000029  0.3333333429999996  0.5217140548665616
	0.6666666870000029  0.3333333429999996  0.0634955699714910	
	0.6238378061607577  0.5940989414527301  0.2330614375419467
	0.3715115645822017  0.3913852574380640  0.7499251076346213
	0.4059010585472700  0.0297388347080183  0.2330614375419467
	0.6086147425619355  0.9801263371441329  0.7499251076346213
	0.9702611952919840  0.3761622238392447  0.2330614375419467
	0.0198736928558696  0.6284884654178010  0.7499251076346213
	0.4970341950086190  0.6419127807282180  0.2504819573050943
	0.4722574174374839  0.3124116086150668  0.7489819492636393
	0.3580872192717823  0.8551214142804011  0.2504819573050943
	0.6875883913849331  0.1598458078224275  0.7489819492636393
	0.1448785557195963  0.5029657759913753  0.2504819573050943
	0.8401541621775773  0.5277425535625101  0.7489819492636393
	0.7465890784344098  0.6974373665162791  0.0877326196589775
	0.2518710331486960  0.3302579489920529  0.9082269759127742
	0.3025626334837207  0.0491516819181211  0.0877326196589775
	0.6697420510079468  0.9216131131566275  0.9082269759127742
	0.9508483470818850  0.2534109515655927  0.0877326196589775
	0.0783869158433784  0.7481289968513065  0.9082269759127742
	0.2751067320564097  0.3631086026138700  0.5737407252403873
	0.7132743878686607  0.6201447197159712  0.4140528338576862
	0.6368913973861298  0.9119981584425314  0.5737407252403873
	0.3798552802840290  0.0931296381526946  0.4140528338576862
	0.0880018705574747  0.7248932979435926  0.5737407252403873
	0.9068703908473114  0.2867256421313418  0.4140528338576862
	0.5441603686812513  0.4191736441937305  0.1764358135764958
	0.4780900223868556  0.5733419996779769  0.7722935441387523
	0.5808263258062671  0.1249867244875282  0.1764358135764958
	0.4266579703220141  0.9047479627088743  0.7722935441387523
	0.8750132455124694  0.4558396313187486  0.1764358135764958
	0.0952520072911234  0.5219099776131445  0.7722935441387523
	-0.0000000000000000  0.0000000000000000  0.2917658511111268
	
\end{verbatim}

\begin{verbatim}
	Pb9Ni(PO4)6O
	1.00000000000000
	9.8154531593033294    0.0000006796980621    0.0000000000000001
	-4.9077271683365957    8.5004314457636099   -0.0000000000000000
	-0.0000000000000000    0.0000000000000000    7.3864048600067536
	Pb   Ni   P    O
	9     1     6    25
	Direct
	0.0023930570875375  0.7740062083079475  0.2599301159193992
	0.9981263570376689  0.2484453590588505  0.7611097394955972
	0.2259937916920525  0.2283868487795829  0.2599301159193992
	0.7515546409411493  0.7496809989788294  0.7611097394955972
	0.7716131662204218  0.9976069569124635  0.2599301159193992
	0.2503190160211756  0.0018736569623321  0.7611097394955972
	0.3333333129999971  0.6666666269999979  0.0083563744589139
	0.3333333129999971  0.6666666269999979  0.4916993729060725
	0.6666666870000029  0.3333333429999996  0.4939887254346352
	0.6666666870000029  0.3333333429999996  0.9888074251107263
	0.6266506161499268  0.5918699199409470  0.2244060791164619
	0.3812271863310681  0.3914807614839344  0.7587825215394234
	0.4081300800590461  0.0347806662089776  0.2244060791164619
	0.6085192385160657  0.9897464548471361  0.7587825215394234
	0.9652193637910250  0.3733494138500756  0.2244060791164619
	0.0102535751528665  0.6187728436689346  0.7587825215394234
	0.4897391594662527  0.6281642336558425  0.2489403253336887
	0.4820546686219311  0.3088652905428651  0.7673862121357911
	0.3718357663441578  0.8615749258104103  0.2489403253336887
	0.6911347094571351  0.1731893770790694  0.7673862121357911
	0.1384250441895871  0.5102608115337415  0.2489403253336887
	0.8268105929209353  0.5179453023780702  0.7673862121357911
	0.7518336521017459  0.7139248061257816  0.0902259672136851
	0.2562082985300035  0.3329554348200832  0.9124466916157625
	0.2860751938742185  0.0379088159759619  0.0902259672136851
	0.6670445651799166  0.9232528927099121  0.9124466916157625
	0.9620912130240441  0.2481663778982495  0.0902259672136851
	0.0767471362900938  0.7437917314699990  0.9124466916157625
	0.2895518375743092  0.3595968242700853  0.5775506773213896
	0.7126740787308330  0.6140434059575414  0.4093799310935214
	0.6404031757299147  0.9299550423042227  0.5775506773213896
	0.3859565940424590  0.0986306427732962  0.4093799310935214
	0.0700449866957832  0.7104481924257007  0.5775506773213896
	0.9013693862267099  0.2873259512691698  0.4093799310935214
	0.5607077621046025  0.4217205963411343  0.1527077813282273
	0.4884167698367849  0.5753387950552064  0.7771070179708647
	0.5782793736588628  0.1389871657634679  0.1527077813282273
	0.4246611749447841  0.9130779147815733  0.7771070179708647
	0.8610128042365367  0.4392922378953977  0.1527077813282273
	0.0869220552184240  0.5115832301632152  0.7771070179708647
	0.0000000000000000 -0.0000000000000000  0.3314289118382008
	
\end{verbatim}

\begin{verbatim}
	Pb9Zn(PO4)6O
	1.00000000000000
	9.8334400401958248    0.0000006480018838    0.0000000000000000
	-4.9167205813333723    8.5160085573994344   -0.0000000000000000
	0.0000000000000000   -0.0000000000000000    7.3839056703763655
	Pb   Zn   P    O
	9     1   6    25
	Direct
	0.0019823826540479  0.7745389833574755  0.2605685968567902
	0.0005492517510040  0.2495094521966610  0.7609281354007629
	0.2254610166425249  0.2274433992965728  0.2605685968567902
	0.7504905478033389  0.7510398005543468  0.7609281354007629
	0.7725566157034323  0.9980176313459532  0.2605685968567902
	0.2489602144456582  0.9994507622489972  0.7609281354007629
	0.3333333129999971  0.6666666269999979  0.0092920421241173
	0.3333333129999971  0.6666666269999979  0.4917023846699586
	0.6666666870000029  0.3333333429999996  0.4879616587832145
	0.6666666870000029  0.3333333429999996  0.9847445792127363	
	0.6259143338516501  0.5933171010070440  0.2252532299041020
	0.3822381133433937  0.3922924244537182  0.7580217383289192
	0.4066828989929561  0.0325972028445967  0.2252532299041020
	0.6077075755462819  0.9899457188896851  0.7580217383289192
	0.9674028271554057  0.3740856961483525  0.2252532299041020
	0.0100543111103175  0.6177619166566158  0.7580217383289192
	0.4910937622588408  0.6322733513223806  0.2499554912148296
	0.4847413266035737  0.3119647511826648  0.7673767206525192
	0.3677266486776189  0.8588204109364665  0.2499554912148296
	0.6880352488173350  0.1727765744209266  0.7673767206525192
	0.1411795590635306  0.5089062087411530  0.2499554912148296
	0.8272233955790783  0.5152586443964203  0.7673767206525192
	0.7513852734009998  0.7147244060386004  0.0905765954240719
	0.2568117601649497  0.3325059740779390  0.9112845865311741
	0.2852755939613921  0.0366608373623969  0.0905765954240719
	0.6674940259220613  0.9243058150869953  0.9112845865311741
	0.9633391916376089  0.2486147565989954  0.0905765954240719
	0.0756942139130106  0.7431882698350526  0.9112845865311741
	0.2907901127615214  0.3596347159378105  0.5766867280958088
	0.7120123182191316  0.6147601473879272  0.4101790717415690
	0.6403652840621893  0.9311554258237027  0.5766867280958088
	0.3852398526120730  0.0972521408312160  0.4101790717415690
	0.0688446031763033  0.7092099172384810  0.5766867280958088
	0.9027478881687901  0.2879877117808713  0.4101790717415690
	0.5600593234989890  0.4238839768242669  0.1541878947238398
	0.4880216222353552  0.5759472461580564  0.7770246166653357
	0.5761159931757306  0.1361753466747224  0.1541878947238398
	0.4240527238419343  0.9120743160772867  0.7770246166653357
	0.8638246233252747  0.4399406765010177  0.1541878947238398
	0.0879256539227108  0.5119783777646448  0.7770246166653357
	0.0000000000000000 -0.0000000000000000  0.3343691085908144
	
\end{verbatim}

\begin{verbatim}
	Pb9Ag(PO4)6O
	1.00000000000000
	9.9406156256295830    0.0000014316309071   -0.0000000000000000
	-4.9703090526946276    8.6088249452360159    0.0000000000000000
	0.0000000000000000    0.0000000000000000    7.3815177070995590
	Pb   Ag   P    O
	9     1     6    25
	Direct
	0.9955447579595498  0.7644045746843522  0.2509353602597195
	0.9949927298849605  0.2573394633408910  0.7433247032096960
	0.2355954253156481  0.2311401832752049  0.2509353602597195
	0.7426605366591089  0.7376532675440730  0.7433247032096960
	0.7688598317247997  0.0044552560404515  0.2509353602597195
	0.2623467474559317  0.0050072841150407  0.7433247032096960
	0.6666666870000029  0.3333333429999996  0.0097937783126133
	0.3333333129999971  0.6666666269999979  0.9965361097111132
	0.3333333129999971  0.6666666269999979  0.4835217433780268
	0.6666666870000029  0.3333333429999996  0.5070448998213292
	0.6224543720111532  0.5948504549204371  0.2525617534221464
	0.3707022294599365  0.3908319789848494  0.7434265012518808
	0.4051495450795629  0.0276038870907134  0.2525617534221464
	0.6091680210151508  0.9798702804750824  0.7434265012518808
	0.9723961429092889  0.3775456579888496  0.2525617534221464
	0.0201297495249200  0.6292978005400662  0.7434265012518808
	0.5033527712042026  0.6535305546234698  0.2412930483724593
	0.4710849401433576  0.3119139273367794  0.7495567696311050
	0.3464694453765302  0.8498222165807333  0.2412930483724593
	0.6880860726632205  0.1591710118065815  0.7495567696311050
	0.1501777534192646  0.4966471997957914  0.2412930483724593
	0.8408289581934234  0.5289150308566359  0.7495567696311050
	0.7183250028016291  0.6312858593131695  0.0733313605171702
	0.2695585902134868  0.3589354917114473  0.9170778206359134
	0.3687141406868311  0.0870391134884577  0.0733313605171702
	0.6410645082885525  0.9106231275020309  0.9170778206359134
	0.9129609155115483  0.2816750271983734  0.0733313605171702
	0.0893769014979748  0.7304414397865158  0.9170778206359134
	0.2548561473055233  0.3330481904549279  0.5805447120393449
	0.7420730616480391  0.6836840845297045  0.4069712237348472
	0.6669518095450727  0.9218079858505874  0.5805447120393449
	0.3163159154702953  0.0583889471183460  0.4069712237348472
	0.0781920431494186  0.7451438826944794  0.5805447120393449
	0.9416110818816598  0.2579269683519635  0.4069712237348472
	0.5370078419806492  0.4163530057402559  0.2864992188809570
	0.4776500244900591  0.5727569085520745  0.7251055610784928
	0.5836469642597415  0.1206548362403932  0.2864992188809570
	0.4272430614479162  0.9048930559379801  0.7251055610784928
	0.8793451337596041  0.4629921580193508  0.2864992188809570
	0.0951069140620176  0.5223499755099412  0.7251055610784928
	0.0000000000000000  0.0000000000000000  0.2254193596757327
	
\end{verbatim}

\begin{verbatim}
	Pb9Au(PO4)6O
	1.00000000000000
	9.9283792820894288    0.0000013290778343   -0.0000000000000001
	-4.9641907921107915    8.5982280121573389   -0.0000000000000000
	0.0000000000000000   -0.0000000000000000    7.4206648327525606
	Pb   Au   P    O
	9     1     6    25
	Direct
	0.9987233766629506  0.7700297440372798  0.2448161471961586
	0.9969578478045801  0.2566428143246935  0.7388207969009422
	0.2299702559627202  0.2286936326256779  0.2448161471961586
	0.7433571856753066  0.7403150344798909  0.7388207969009422
	0.7713063823743271  0.0012766373370507  0.2448161471961586
	0.2596849805201143  0.0030421661954207  0.7388207969009422
	0.6666666870000029  0.3333333429999996  0.0135593361795980
	0.3333333129999971  0.6666666269999979  0.9968935070071225
	0.3333333129999971  0.6666666269999979  0.4811144980220491
	0.6666666870000029  0.3333333429999996  0.5126567045891999
	0.6240235712206006  0.5957989341703960  0.2567762672942094
	0.3725888757357917  0.3916743705957365  0.7421386506787645
	0.4042010658296044  0.0282246070502167  0.2567762672942094
	0.6083256294042633  0.9809145351400576  0.7421386506787645
	0.9717754229497860  0.3759764587794018  0.2567762672942094
	0.0190854948599448  0.6274111542642108  0.7421386506787645
	0.4984971284963468  0.6460404521305081  0.2409835588602634
	0.4733875116231799  0.3117224796945201  0.7500784349250454
	0.3539595478694922  0.8524566763658462  0.2409835588602634
	0.6882775203054801  0.1616650309286632  0.7500784349250454
	0.1475432936341515  0.5015028425036472  0.2409835588602634
	0.8383349390713415  0.5266124593768144  0.7500784349250454
	0.7157264128121672  0.6264516885192809  0.0762573986474450
	0.2765938771727317  0.3645903707883150  0.9178508294109737
	0.3735483114807189  0.0892746942928838  0.0762573986474450
	0.6354096292116850  0.9120035353844084  0.9178508294109737
	0.9107253347071221  0.2842736171878352  0.0762573986474450
	0.0879964936155975  0.7234061528272706  0.9178508294109737
	0.2537416035714394  0.3286511177541609  0.5839074550456410
	0.7447503840873385  0.6950959851918572  0.4051645076649013
	0.6713488822458390  0.9250905148172633  0.5839074550456410
	0.3049040148081429  0.0496543688954859  0.4051645076649013
	0.0749095141827429  0.7462584264285628  0.5839074550456410
	0.9503456601045199  0.2552496459126570  0.4051645076649013
	0.5439737144232027  0.4184961225329676  0.3021575567533752
	0.4798450214250294  0.5729385081852881  0.7170648712999956
	0.5815038474670371  0.1254775918902420  0.3021575567533752
	0.4270614618147026  0.9069064532397367  0.7170648712999956
	0.8745223781097555  0.4560262855767976  0.3021575567533752
	0.0930935167602609  0.5201549785749704  0.7170648712999956
	0.0000000000000000  0.0000000000000000  0.2019265201688183
	
\end{verbatim}
